\documentclass[aps,pra,twocolumn,showpacs,preprintnumbers,amsmath,amssymb,superscriptaddress,nofootinbib,balancelastpage,showkeys]{revtex4}

\usepackage{amsmath,amssymb}
\usepackage[latin1]{inputenc}
\usepackage{dcolumn}
\usepackage{bm}
\usepackage{graphicx}
\usepackage{dcolumn}
\usepackage{bm}
\usepackage{upgreek}

\begin{document}


\title{Controlling passively-quenched single photon detectors by bright light}

\author{Vadim Makarov}
\email{makarov@vad1.com}
\affiliation{Department of Electronics and Telecommunications, Norwegian University of Science and Technology, NO-7491 Trondheim, Norway}
\affiliation{Department of Physics, Pohang University of Science and Technology, Pohang 790-784, Korea}

\date{April~17, 2009}

\begin{abstract}
Single photon detectors based on passively-quenched avalanche photodiodes can be temporarily blinded by relatively bright light, of intensity less than a nanowatt. I describe a bright-light regime suitable for attacking a quantum key distribution system containing such detectors. In this regime, all single photon detectors in the receiver Bob are uniformly blinded by continuous illumination coming from the eavesdropper Eve. When Eve needs a certain detector in Bob to produce a click, she modifies polarization (or other parameter used to encode quantum states) of the light she sends to Bob such that the target detector stops receiving light while the other detector(s) continue to be illuminated. The target detector regains single photon sensitivity and, when Eve modifies the polarization again, produces a single click. Thus, Eve has full control of Bob and can do a successful intercept-resend attack.
To check the feasibility of the attack, 3 different models of passively-quenched detectors have been tested. In the experiment, I have simulated the intensity diagrams the detectors would receive in a real quantum key distribution system under attack. Control parameters and side effects are considered. It appears that the attack could be practically possible.
\end{abstract}

\maketitle

\section{\label{sec:intro}Introduction}

Quantum key distribution (QKD) is a technique that allows remote parties to grow shared secret random key material at a steady rate, using an insecure optical communication channel and an authenticated classical communication channel \cite{bennett1992,quantum-cryptography-reviews}. Since a tabletop demonstration nineteen years ago \cite{bennett1992}, QKD has progressed to commercial devices working over tens of kilometers of optical fiber \cite{commercial-QKD-systems} and many long-distance experiments. Key transmission over more than a hundred kilometers of fiber \cite{over-100km-non-decoy,over-100km-decoy}, 23~km and 144~km of free space \cite{PhysRevLett-98-p010504,NaturePhysics-adv-nphys629,Nature-419-p450-and-ProcSPIE-4917-p25} has been demonstrated. As QKD enters commercial market, it becomes increasingly important to verify the actual level of security in its implementations, and search for possible loopholes.

QKD has been proven to be unconditionally secure for certain models of equipment that include most common imperfections of components \cite{unconditional-proofs}. However, it remains a challenge to build a system that is demonstrably in strict accordance with the model in the security proof. Discovering and patching loopholes and imperfections of components is an ongoing process \cite{JModOpt-48-p2039,large-pulse-attack,JModOpt-52-p691,detector-efficiency-mismatch,arXiv-0809-3408-quant-ph}. Once found, such imperfection affecting security can either be integrated into the unconditional proof, neutralized by a specific coutermeasure, or avoided through a rational choice of components, optical scheme and QKD protocol. 

When treating security of QKD, we follow Kerckhoffs' principle: ``The system must not be required to be secret, and it must be able to fall into enemy's hands without causing inconvenience'' \cite{JdesSciencesMilitaires-IX-jan1883-5}. This principle, embraced in the classical cryptography since the 19th century, means Eve is assumed to know everything about Alice's and Bob's equipment. Thus, Eve can fully exploit every imperfection that exists in legitimate parties' hardware and software. Although it's tempting to assume Eve might not know the type of equipment or its exact parameters, the history of cryptography shows she will eventually find this out. In QKD, practical ways of measuring unobtrusively equipment parameters of a running cryptosystem may exist as well \cite{JModOpt-52-p691}.

In this paper, I report an imperfection found in single photon detectors (SPDs) of one particular type, namely those based on passively-quenched avalanche photodiodes (APDs). This particular type of SPD is used in probably about 10\% of all QKD implementations reported up to the date. Since the passive quenching is most suited for silicon APDs, the majority of the possibly affected systems are free-space QKD experiments doing optical transmission in the 500--900~nm wavelength range; they are listed in Sec.~\ref{sec:countermeasures}. The current commercial devices working at longer telecommunication wavelengths \cite{commercial-QKD-systems} are {\it not} affected by this paticular vulnerability, because they use another type of SPD, a gated APD.

\section{\label{sec:controlling-single-detector}Blinding and controlling a passively-quenched single photon detector}

Passive quenching is the oldest and simplest possible circuit design in SPDs based on APDs \cite{JApplPhys-35-p1370-and-JApplPhys-36-p3123,ApplOpt-35-p1956}.
\begin{figure*}
\includegraphics[height=82mm]{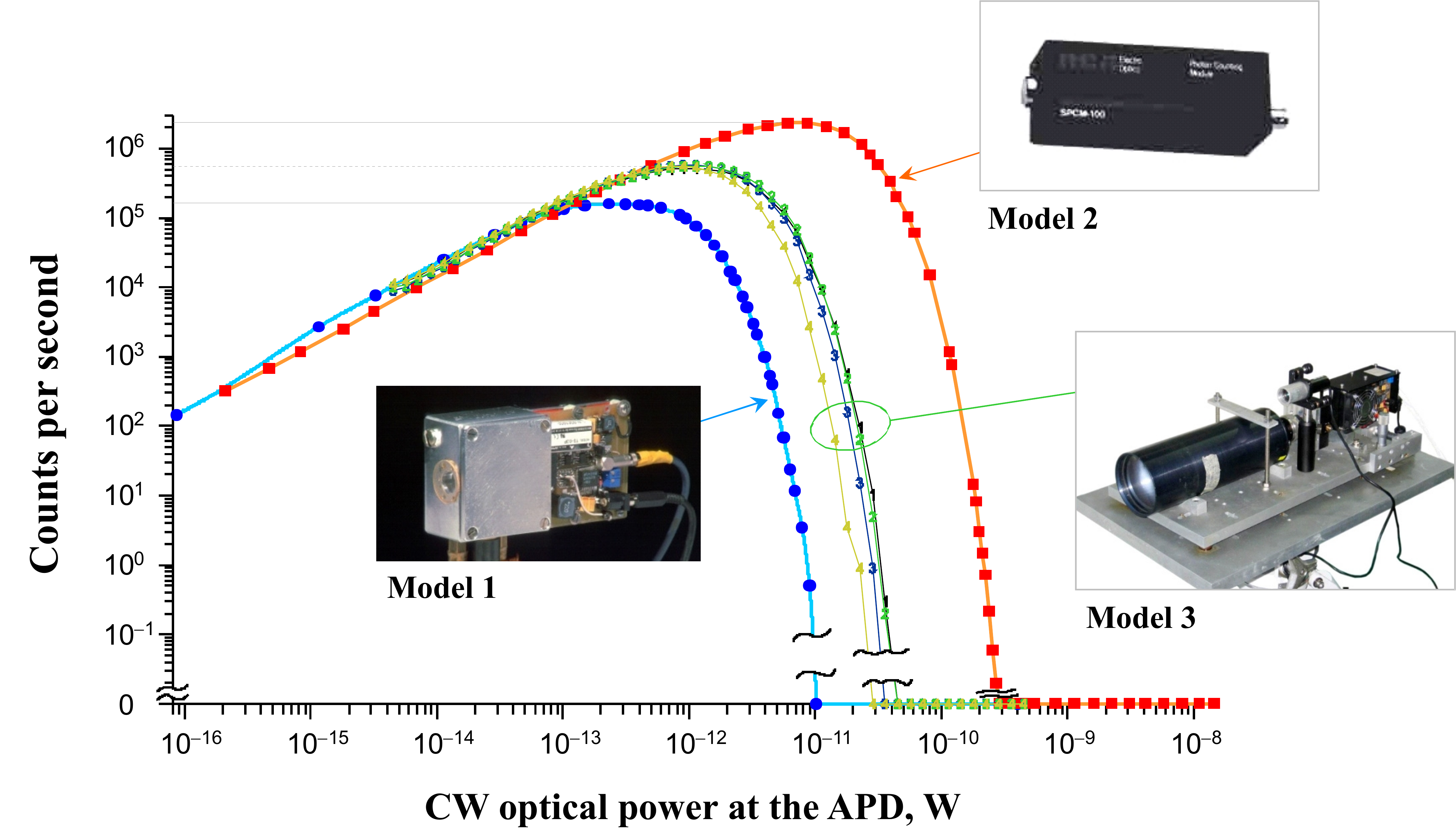}
\caption{\label{fig:saturc}Detector saturation curves. Model~1: do-it-yourself design by C.~Kurtsiefer and his coworkers (currently used in several laboratories; the sample tested was assembled at the Laboratory of spontaneous parametric down-conversion at the Moscow State University). Model~2: EG\&G SPCM-200-PQ (industrially produced in the 1990s). Model~3: four detectors used in Bob in a daylight free-space QKD system \cite{arXiv-0812-1880-quant-ph} (entire Bob is pictured in the inset; curves for model~3 reprinted from \cite{Qin-upcoming}). The dark count rate is around 100 counts per second (cps) for model~1, around 50~cps for model~2, and in the 900--2100~cps range for model~3.}
\end{figure*}
Beyond the useful photon counting rate range, passively-quenched SPDs exhibit saturation and blinding behavior. Fig.~\ref{fig:saturc}
illustrates this on the example of three different SPD models I have tested. Up to a certain point different for each SPD model, their count rate increases approximately linearly with intensity of CW illumination. At higher input light intensities, the count rate saturates, reaches the peak value different for each model, and begins to drop. It drops to {\em exactly} zero at 10~pW input power (at 820~nm wavelength) for model~1, at 280~pW (at 780~nm) for model~2, and at intermediate power values for the four tested detectors of model~3. The shapes of the saturation curves are similar for all the tested detector models. This suggests that the saturation and blinding is generic to the passively-quenched detector design. For the rest of this paper (except Sec.\ \ref{sec:detector-model-2}), characteristics of the model~1 are given in all examples, while the models~2 and 3 are implied to exhibit the same behavior with different values of parameters.

\begin{figure}
\includegraphics[width=87mm]{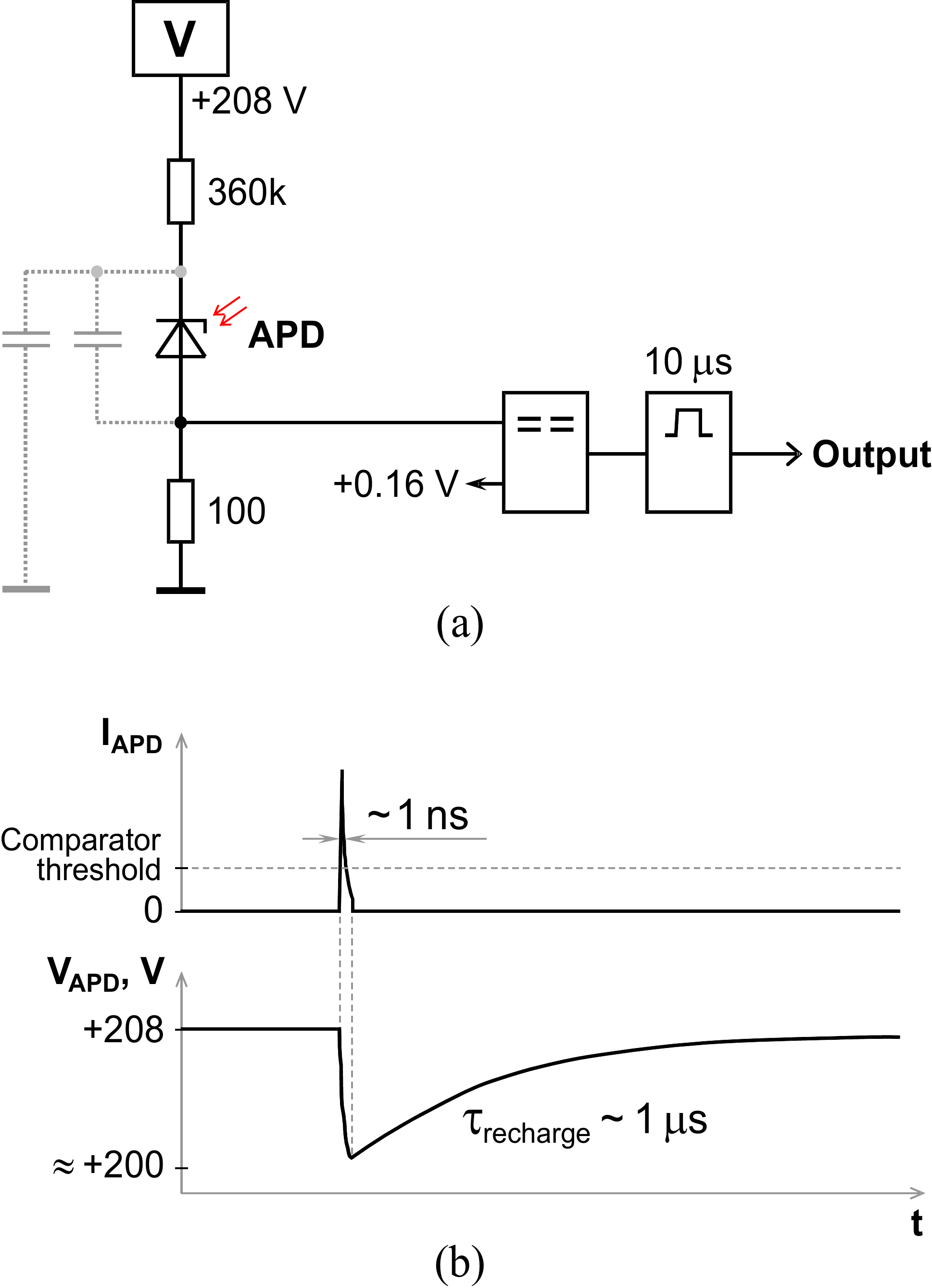}
\caption{\label{fig:detector}Detector model~1: ~(a)~equivalent circuit diagram; ~(b)~current through the APD and voltage at the APD during an avalanche and subsequent recharge.}
\end{figure}

To explain the blinding behavior, let's consider the circuit diagram of the detector model~1 (Fig.~\ref{fig:detector}).
\begin{figure}
\includegraphics[width=66mm]{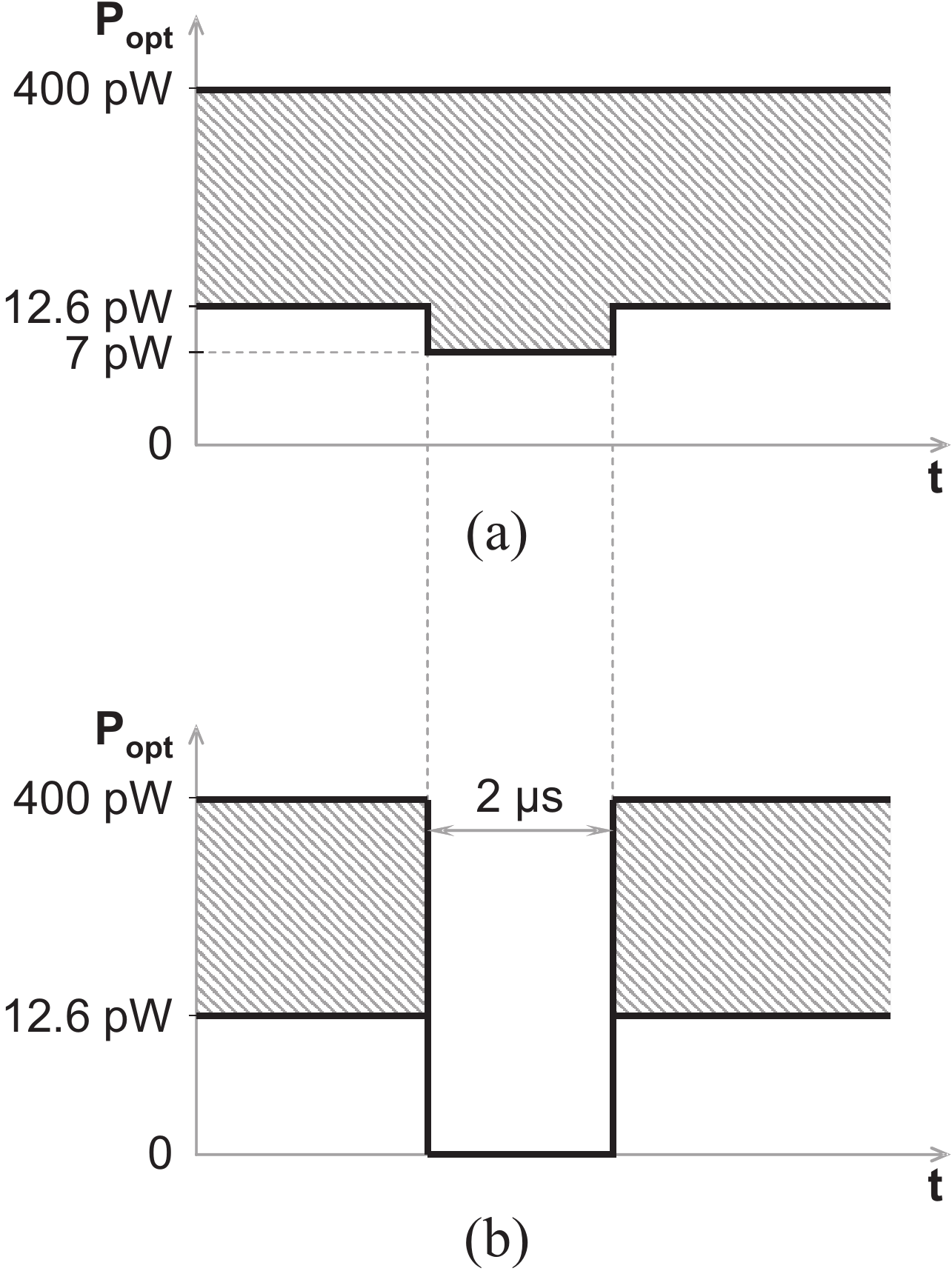}
\caption{\label{fig:single-det-control-diagrams}Control diagrams for detector model~1: ~(a)~input intensity diagram that keeps the detector completely blinded at all times (no output pulses); ~(b)~input intensity diagram that produces a single output pulse with probability greater than 0.8, right after the end of the ${{\rm 2~\upmu s}}$ gap. The actual input intensity on both diagrams may take any shape within the hatched area.}
\end{figure}
The Si APD (PerkinElmer C30902S) is biased six to ten volts above its breakdown voltage from a high-voltage source via a 360~k$\Omega$ resistor. The circuit works thanks to the presence of two stray capacitances of the order of 1~pF each, shown in the circuit diagram. When there is no current flowing through the APD, both capacitances are charged to the bias voltage. During an avalanche, they quickly discharge through the APD, producing a short current pulse. The discharge current of the leftmost capacitance is converted into voltage at a 100~$\Omega$ resistor, and this voltage is sensed by a fast ECL comparator (MC100EL16). The short output pulse of the comparator is widened to about ${{\rm 10~\upmu s}}$ by a non-retriggerable monostable multivibrator. The current pulse produced during the avalanche is on the order of 1~ns wide. When the voltage at the APD drops sufficiently close to the breakdown voltage, the avalanche quenches. The capacitances are subsequently slowly recharged through the bias resistor, with a recharge time constant on the order of ${{\rm 1~\upmu s}}$.

Until the capacitances recharge to a certain threshold voltage, which in our detector sample takes about ${{\rm 1~\upmu s}}$, the detector has no single photon sensitivity. (After ${{\rm 1~\upmu s}}$, it increases its quantum efficiency gradually as the voltage continues to rise.) However, a photon coming during the first microsecond may still cause an avalanche with a smaller peak current, not reaching the comparator threshold \cite{ApplOpt-35-p1956}. Such small avalanches reset the voltage and can keep the detector blinded indefinitely if they occur often enough. This is the primary blinding mechanism in the passively-quenched detectors. Additionally, heating of the APD chip can contribute to the blinding. At 10~pW input optical power, the average electrical power dissipated in the APD is measured to be 5.7~mW. PerkinElmer C30902S APD is reported to have a high thermal resistance between the chip and the package \cite{ApplOpt-32-p3894}. The measured electrical power may raise the chip temperature by several degrees. This rise in temperature would increase the breakdown voltage by several volts, which could be a contributing factor to the blinding.

In applications of SPDs, the non-linearity of the response is undesirable \cite{ApplOpt-35-p1956}. Typically a detector is considered usable only in the mostly linear portion of its characteristic, located to the left of the saturation peak in Fig.~\ref{fig:saturc}. Detectors are never used beyond their saturation point. The following might be the first ``useful application'' I have found for the beyond-saturation regime. In doing an attack against a QKD system, Eve may blind Bob's SPDs by delivering constant illumination higher than 10~pW to each of them. However, by introducing a gap in which the intensity of illumination drops to zero at one of Bob's SPDs, she may induce an output pulse at that SPD.

Let's first consider how Eve can control a single SPD. Experimental tests made on the detector model~1 have demonstrated that the control diagrams shown in Fig.~\ref{fig:single-det-control-diagrams}
can be used. When the power of input illumination $P_{\rm opt}$ stays within the range depicted in the diagram (a), the SPD is kept blinded. However, in the diagram (b), after the light is switched off, the capacitances in the SPD have time to recharge and it becomes sensitive to single photons. When the light is switched on ${{\rm 2~\upmu s}}$ later, the SPD produces a single photon count with probability greater than 0.8 (or no click in the remaining fraction of the cases), and after that becomes blinded again. I have only tested power values up to 400~pW with this detector model; however the upper border of the power range could probably be extended much higher than 400~pW without causing any new effects. Experimental tests of the detectors are treated in more detail later on, in Sec.~\ref{sec:quality-of-control}.

\section{\label{sec:attack-against-QKD-system}Proposed attack against QKD system}

\begin{figure}
\includegraphics[width=78.7mm]{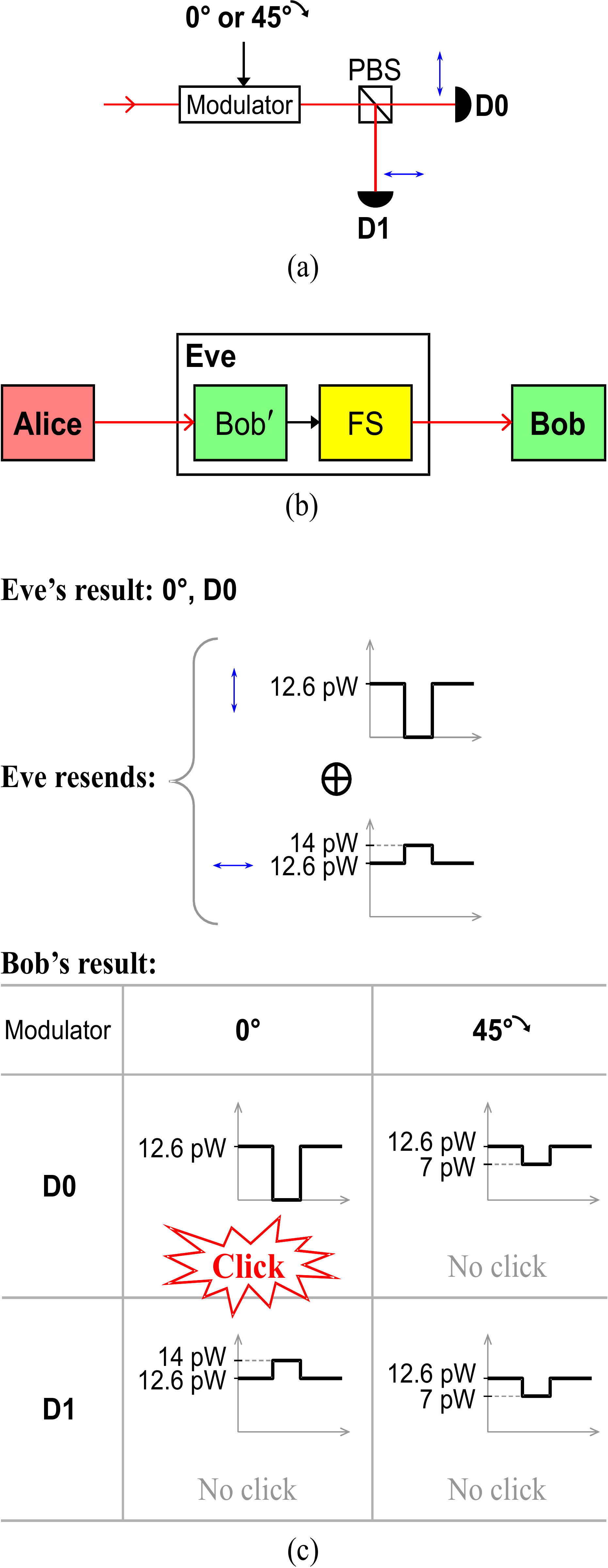}
\caption{\label{fig:faked-states-attack}Proposed attack against a QKD system that uses two model~1 detectors: ~(a)~equivalent optical scheme of Bob's setup; ~(b)~scheme of the faked-state attack; ~(c)~faked state sent by Eve in case of her detection in the $0\textdegree$ basis with the result D0, and the intensity diagrams that result at Bob's detectors for his two possible basis choices. Optical losses in components are neglected; in the presence of losses, Eve should send a proportionally brighter faked state.}
\end{figure}

With the detector control method described above, Eve can attack a complete QKD system. In a QKD system, Bob has several detectors and/or makes a choice of detection basis. Eve needs a way to cause a click in a specific detector in a specific basis of her choice, without causing a click in the other detector(s) nor in a different basis. I initially explain the attack on an example of a system with polarization coding and active basis choice at Bob that runs the Bennett-Brassard 1984 (BB84) protocol \cite{BB84-orig,bennett1992}. In such a system, input light at Bob first passes through a modulator that, at Bob's random choice, either does nothing or rotates any input polarization state {45\textdegree} clockwise, thus setting one of the two possible detection bases (Fig.~\ref{fig:faked-states-attack}(a)).
After the modulator, light is split at a polarizing beamsplitter PBS. The vertical component of polarization goes to the detector D0 and horizontal component goes to the detector D1.

Eve runs an intercept-resend attack (faked-state attack \cite{JModOpt-52-p691}) against this system. In the faked-state attack, she blocks the light between Alice and Bob completely (Fig.~\ref{fig:faked-states-attack}(b)). Eve uses a replica of Bob's setup Bob$^\prime$ to detect Alice's quantum state, choosing the detection basis at random. Then, Eve forces Bob to make a click in her basis only and with the same bit value as she has just detected. (This is the difference between the conventional intercept-resend attack \cite{bennett1992} and the faked-state attack \cite{JModOpt-52-p691}: in the latter, the basis and bit value of Bob's detector click is always the same as Eve's, thus the attack does {\it not} cause errors in the sifted key and eavesdropping is not detected.) Eve forces a click in the selected basis with the specific bit value by sending to Bob a specially crafted light state called {\it faked state,} using her faked state generator FS. The faked state exploits technological imperfections in Bob to achieve its goal. In the present study, it will exploit detector controllability.

Let's suppose for certainty that Eve has detected Alice's quantum state in the {0\textdegree} basis and registered a click in her D0 detector. She now has to form and send to Bob a faked state. The faked state should cause a click in Bob's D0 detector in the case Bob chooses the {0\textdegree} basis (the same basis as Eve has used), and cause no clicks in either of Bob's detectors in the case Bob chooses the {45\textdegree} basis (not the basis Eve has used). The faked state that reaches this goal consists of an incoherent mixture of vertical and horizontal polarization components, with an intensity diagram for each polarization component as shown in the upper half of Fig.~\ref{fig:faked-states-attack}(c). The lower half of Fig.~\ref{fig:faked-states-attack}(c) shows what happens to this faked state in Bob's setup. If Bob chooses the {0\textdegree} basis, his modulator does nothing and the two polarization components of the faked state are split each to its own detector. The intensity diagram of the vertical polarization component causes a click in D0 with probability greater than 0.8 (for the gap width of ${{\rm 2~\upmu s}}$). The intensity diagram of the horizontal polarization component keeps D1 blinded. If Bob, however, chooses the {45\textdegree} basis, each polarization component is rotated {45\textdegree} and is split equally at the polarizing beamsplitter. The halves of the two polarization components sum at each detector, resulting in identical intensity diagrams that keep both detectors blinded. Eve's three other possible bit-basis detection results are treated symmetrically. Thus, my faked-state attack succeeds.

The reader may notice that the probability of a faked state sent by Eve to cause a click at Bob is $\frac{0.8}{2}=0.4$. Many realistic Bobs have overall photon detection efficiency less than 40\%, mainly due to limited quantum efficiency of the APDs. For these Bobs, Eve can mimic their detection rate before the attack, provided she uses ideal SPDs with 100\% quantum efficiency and zero-loss optics in Bob$^\prime$. However, I want my attack not only be possible {\it in principle,} but implementable in practice, today. For that, Eve cannot uses non-existent ideal detectors. It would also be impractical for her to use exotic high quantum efficiency detectors working at cryogenic temperatures. Most practical for Eve would be to use a copy of actual Bob's setup for her Bob$^\prime$, maybe with limited improvements. In this situation, Bob will observe loss of detection efficiency under attack, which equivalently appears to him as a sudden 4~dB additional loss in the line, and may trigger a security alarm. However, this would be a solvable problem for Eve in most of the real situations, because loss in the transmission line between Alice and Bob almost always exceeds 4~dB. Eve may place her detection unit closer to Alice. Thus, she excludes the loss in the length of the line between her detection unit and Bob. This compensates the reduced ``detection efficiency'' of her faked states at Bob. Eve could also try to improve quantum efficiency of her detectors and reduce losses in Bob$^\prime$ comparing to those of Bob's setup. In free-space QKD, the losses Eve could try to reduce would include the coupling loss from Alice's free-space beam into the receiver telescope.

My attack is applicable to different protocols and schemes, when they use vulnerable detectors. The attack clearly applies to schemes with passive basis choice at Bob \cite{ElectronicsLett-37-p512,NewJPhys-4-p43,PhysRevLett-89-p187901,Nature-419-p450-and-ProcSPIE-4917-p25,NewJPhys-6-p92,OptExpress-12-p3865,OptExpress-13-p202,PhysRevLett-94-p150501,ProcQuantumElectronicsConference-2005-EQEC05-305,ApplPhysLett-89-p101122,arXiv-0812-1880-quant-ph,PhysRevLett-98-p010504,NaturePhysics-adv-nphys629,OptExpress-16-16840}. For these schemes, Eve should double the intensity of her faked states. The random basis choice by Bob is removed: Eve always gets to choose the basis for him. In the case of the BB84 protocol, the four cells in the table in Fig.~\ref{fig:faked-states-attack}(c) represent the intensity diagrams at the four Bob's detectors. This is the case described in the abstract of this paper.

Besides polarization, another coding widely used in QKD is phase coding \cite{PhysRevLett-68-3121,Townsend-phase-coding}. If a scheme with phase coding employs vulnerable detectors, this attack can be run against it. For phase coding Eve shall, instead of the polarization components shown in Fig.~\ref{fig:faked-states-attack}(c), use components of faked state with 0 and $\pi$ phase differences between the arms of the interferometer. The attack will also work on systems using the Scarani-Acin-Ribordy-Gisin 2004 (SARG04) protocol \cite{SARG-protocol} and most of the decoy-state protocols \cite{decoy-states,PhysRevA-75-050305R,PhysRevLett-98-p010504,over-100km-decoy}, as long as Bob is using passively-quenched detectors. The decoy-state protocols referenced above do not help the legitimate users against this attack, because Eve does not measure photon number. She detects Alice's states with a faithful replica of Bob's setup and then simply forces her detection results onto Bob as transparently as she can. At last, this attack is also applicable to the Bennett 1992 (B92) protocol \cite{PhysRevLett-68-3121,PhysRevLett-81-p3283,PhysRevLett-84-p5652,OptExpress-12-p2011,ApplPhysLett-89-p191121,IEEEJQuantumElectron-40-p900,OptExpress-13-p3015,IEEEJQuantumElectron-43-p130}, to the Ekert protocol \cite{PhysRevLett-67-661,PhysRevA-78-020301R}, to the six-state protocol \cite{six-state-protocol} and, under certain conditions, to secret sharing schemes \cite{secret-sharing}.

For a practical implementation of the attack, it is important to consider all side effects it causes, and how to work them around so that Alice and Bob are not alarmed. One side effect, the less-than-unity ``detection efficiency'' of the faked states at Bob, has been discussed above. Another side effect is the replacement of dark counts of Bob's detectors with dark counts of Eve's detectors. During the attack, Eve keeps Bob's detectors blinded when she is not sending faked states. Thus, they do not produce spontaneous counts. Instead, Eve has dark counts in her detectors which she cannot distinguish from Alice's photons. She passes them on to Bob as faked states. Eve's detectors may have a lower ratio of dark counts to photon counts than Bob's. Eve is certainly allowed to achieve this in practice, either by using better detectors or by placing them closer to Alice (which she may have to do anyway). This may cause an overall reduction in the quantum bit error rate (QBER) experienced by Alice and Bob, and be noticed by them. If this becomes a problem, Eve can emulate additional dark counts by sending random faked states to Bob at random times. Similarly, optical imperfections at Bob that originally contributed to the QBER get replaced by the optical imperfections in Eve's copy of Bob's setup. (However, the optical imperfections at Bob may still make some contribution to the QBER through timing side effects during the attack, as will be shown in the next section.)

Side effects may arise when Eve begins and ends the attack. When she goes into the control mode by switching on the constant illumination, Bob's detectors will each produce a single click. These initial clicks at the beginning of the attack may register as one or more error bits in the key. However, this should not be a problem as long as Eve does not switch in and out of the attack mode too frequently. At the end of the attack, when Eve switches off the illumination, no extra clicks are produced except for maybe afterpulses with slightly increased probability than normal. Thus, at least in principle, Eve can begin and end this attack on a running quantum cryptolink.

Another side effect is the additional delay in the quantum channel caused by Eve. The major component of this delay is the gap width in the faked state. Eve begins forming the faked state immediately after detecting Alice's quantum state. However, the actual click at Bob occurs at the end of the gap, which comes ${{\rm 2~\upmu s}}$ later. Thankfully, the time on Alice's and Bob's clocks is not authenticated in the QKD protocol. Many of the possibly affected QKD systems (listed in Sec.~\ref{sec:countermeasures}) measure the time difference by the time of arrival of quantum states to Bob. In these systems, the additional ${{\rm 2~\upmu s}}$ delay will easily be absorbed by the time synchronization algorithm. In case the delay ever becomes a problem for Eve, she may try a slightly different tactics. Eve could begin sending to Bob a faked state for a particular bit-basis combination before she actually detects it. Then, when she detects Alice's quantum state in this bit-basis combination, she instantly ends the gap and finishes the faked state. As will be shown in the next section, the gap in the faked state can be of variable width, so this tactics might work.

Finally, Eve must take into account two practical limitations of the hardware. One limitation is a finite extinction ratio of Bob's PBS, as well as Eve's finite precision in forming polarized light with exact parameters of polarization. The resulting imperfect splitting of the two faked state components at Bob's PBS leads to non-zero optical power in the gap on the control intensity diagram of the target detector. Another limitation is the time distribution of detector counts induced immediately after the end of the gap. This time distributon has a non-negligible width. These two limitations and their effects on the attack are considered in the next section.

\section{\label{sec:quality-of-control}Detector tests}

In this section, I mainly consider time distribution of clicks induced by faked states. Many of the possibly affected QKD systems register timing of detector output pulses with sub-nanosecond precision. The width of Bob's time bin in which clicks are accepted as belonging to a particular Alice's qubit can be on the order of a nanosecond. Ideally, Eve's faked state should induce a click with sub-nanosecond time precision, to target the qubit time bin. However, as the tests show, the actual time distribution of the induced clicks is much wider.

The experimental tests of three different detector models are reported below.

\subsection{\label{sec:detector-model-1}Detector model~1}

This detector model is based on a solder-it-yourself printed circuit board developed by C.~Kurtsiefer and his coworkers. Being a low-cost, simple and compact design, it is used in several laboratories around the world. The equivalent diagram of the signal part of the circuit is shown in Fig.~\ref{fig:detector}(a). The particular sample I have tested features multivibrator pulse duration of about ${{\rm 10~\upmu s}}$, while it is usually made orders of magnitude shorter in this circuit.

\begin{figure}
\includegraphics[width=57mm]{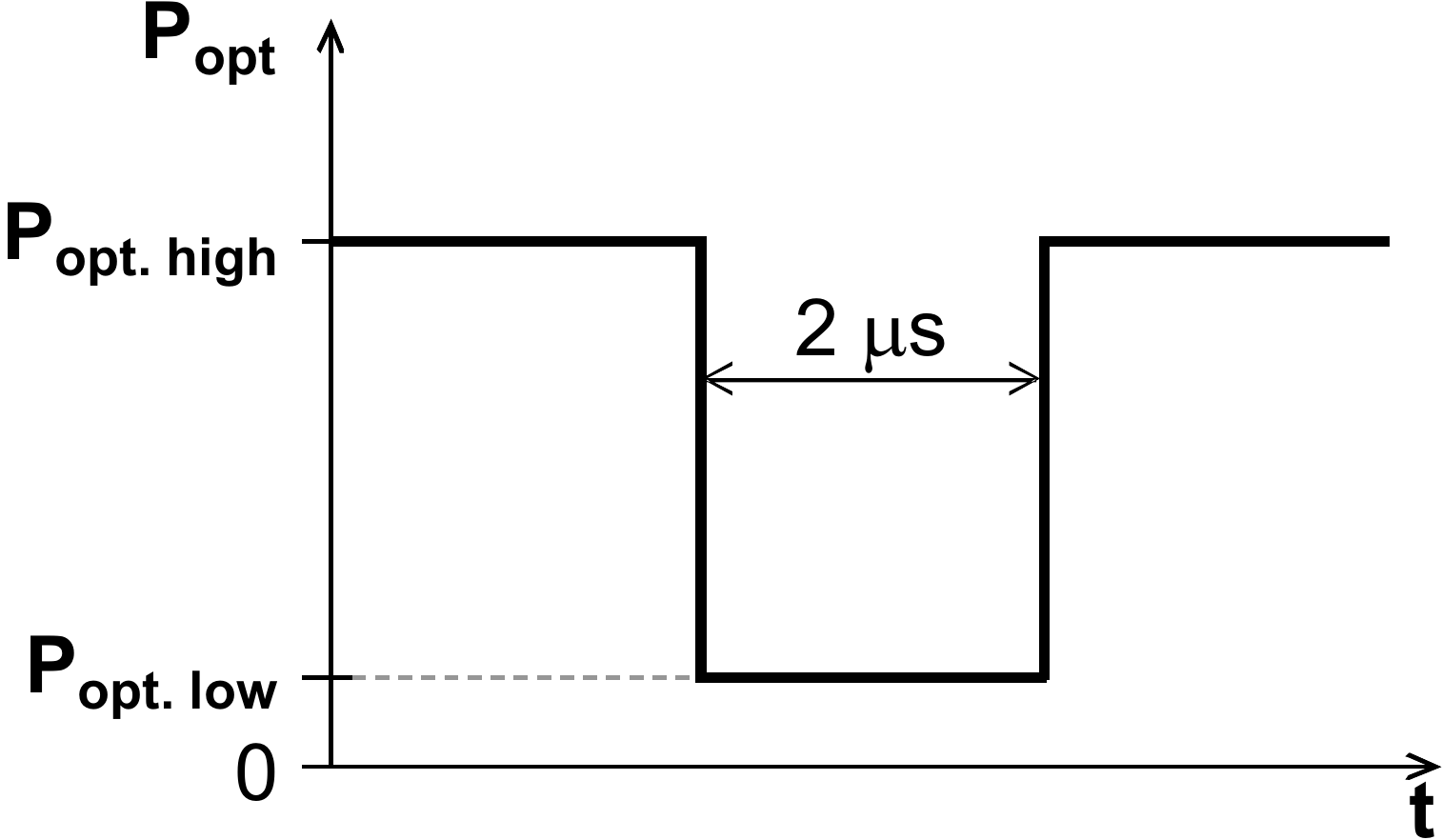}
\caption{\label{fig:detmodel1-control-diagram}Detector model~1. Control intensity diagram during testing. It was applied at 1~kHz repetition rate.}
\end{figure}

The detector has been tested under input illumination time diagram shown in Fig.~\ref{fig:detmodel1-control-diagram}.
Laser illumination at 820~nm wavelength was applied uniformly over the entire photosensitive area of the APD 0.5~mm in diameter.\footnote{A detailed description of the testing setup can be found in the first version of this article, arXiv:0707.3987v1 [quant-ph].} The optical power values $P_{\rm opt}$ are calculated as the total power impinging on the photosensitive area. I have tested the detector at both zero and non-zero power level in the gap $P_{\rm opt.~low}$.

\begin{figure}[t]
\includegraphics[width=87mm]{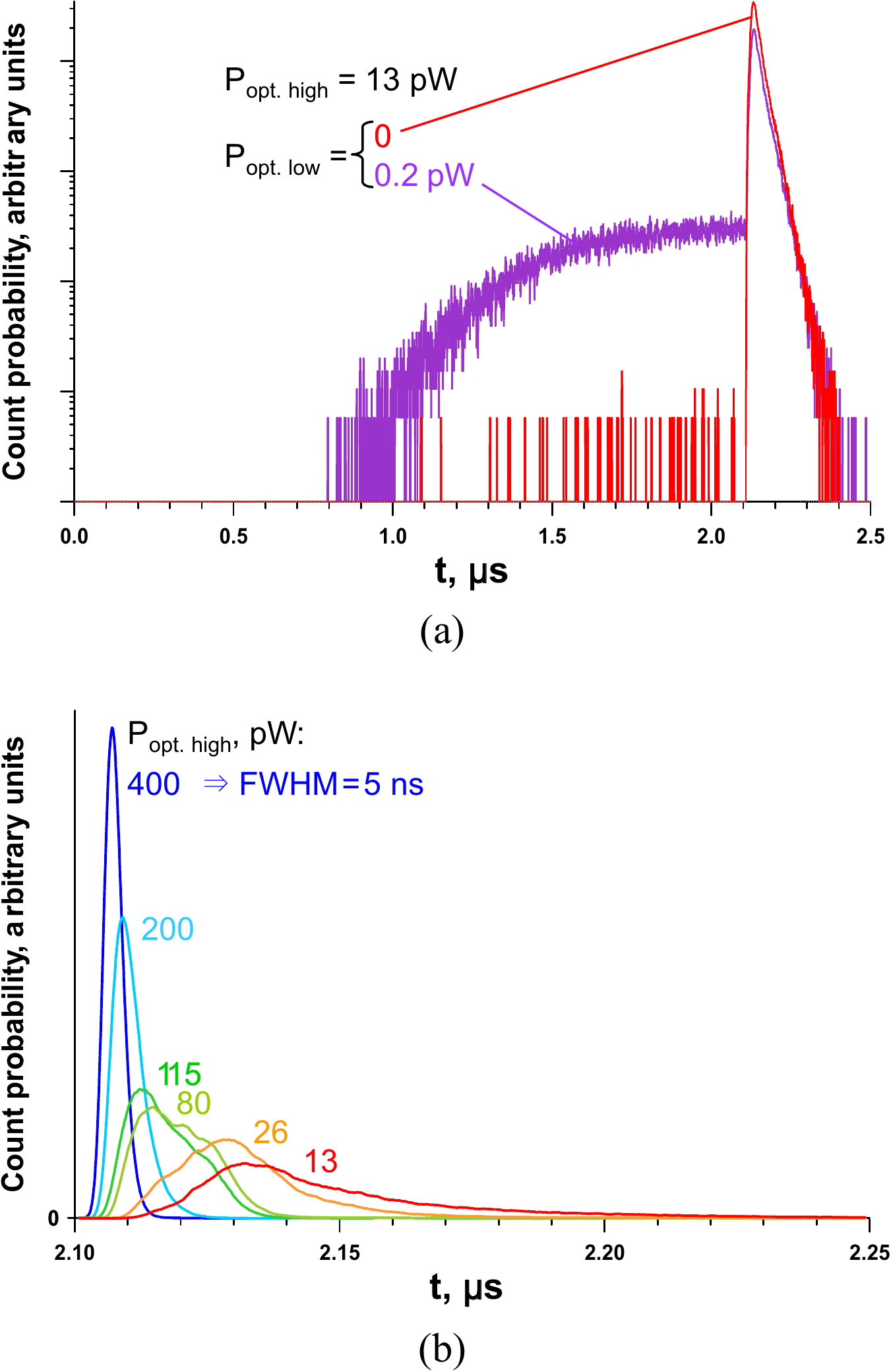}
\caption{\label{fig:timedistr}Detector model~1. Time distributions of pulse's leading edge at the detector output, when Eve controls it by constant illumination with an approximately ${{\rm 2~\upmu s}}$ wide gap: ~(a)~time distributions for zero and non-zero $P_{\rm opt.~low}$ (at $P_{\rm opt.~high}={\rm 13~pW}$); ~(b)~time distributions in the main response peak for a range of $P_{\rm opt.~high}$ values (at $P_{\rm opt.~low}=0$). On the charts, $t=0$ approximately corresponds to the start of the gap in illumination; the main response peak begins just after the end of the gap.}
\end{figure}

Figure~\ref{fig:timedistr}(a)
shows a typical time distribution of the SPD output pulses, and what effects non-zero power in the gap has on this time distribution. During approximately the first ${{\rm 1~\upmu s}}$ of the gap, the SPD does not produce output pulses at all. After ${{\rm 1~\upmu s}}$, some premature output pulses appear. When there is no illumination in the gap ($P_{\rm opt.~low}=0$), the average rate of these pulses is, at the parameters for which the chart is plotted, between three and four times the normal dark count rate (of about 100~cps). After the end of the gap, there is a main response peak of a certain width. Non-zero illumination in the gap causes two effects. Firstly, the probability of premature output pulses greatly increases, as can be seen on the $P_{\rm opt.~low}={\rm0.2~pW}$ curve. Secondly, the probability of output pulses in the main response peak decreases.

The width of the main response peak can be reduced by increasing $P_{\rm opt.~high}$, as shown in Fig.~\ref{fig:timedistr}(b). 5~ns full width at half maximum (FWHM), or 10~ns width as measured near the base of the peak at the 2\% magnitude level, has been achieved at $P_{\rm opt.~high}={\rm 400~pW}$. The width could likely be decreased further at higher levels of $P_{\rm opt.~high}$; however, I did not test beyond 400~pW with this detector model. The detector response in the main peak is a single-photon click, as suggested by the exponentially decaying tail of the time distribution and by an estimate of the number of photons impinging on the APD in a unit of time. It is possible that multi-photon effects influence the time distribution at higher levels of $P_{\rm opt.~high}$; however there was no practical way of testing this. I have not investigated which effect is responsible for the gradual rising edge of the main response peak in this test.

As you see, the total width of the time distribution, including the premature clicks, is more than a microsecond. The practical significance of this wide time distribution varies a lot depending on how Bob treats clicks falling outside his qubit time bin (which is always much narrower than a microsecond). If all or most of these clicks are simply disregarded by him, this is not much of a problem for Eve. In this case she only faces an additional reduction in the ``detection efficiency'' of her faked states at Bob, which could be compensated as discussed in the previous section. If, however, clicks registered by Bob outside the proper qubit time bin contribute to the QBER (by falling into adjacent qubit time bins) or trigger an alarm condition, then Eve faces more stringent requirements. How these clicks are actually treated depends on implementation details and algorithms in each particular QKD system under attack, which I do not consider here. The relevant implementation details are usually not reported in papers to the required extent, so experimenting with each QKD system will be necessary.

We can still estimate how bad this problem can be by considering one of the worst possible cases for Eve. While the width of the main response peak can be reduced by increasing $P_{\rm opt.~high}$, the premature counts in the gap are always distributed over a wide time span. I assume that all these premature counts fall into wrong qubit time bins at Bob. This can happen in a high-speed QKD system with qubit time bins following each other with no gaps between them, and passive basis choice at Bob. In the BB84 protocol, a count falling into a wrong qubit bin has 25\% chance of causing an error in the sifted key (a combination of 50\% chance of being in a compatible basis and 50\% chance of having a wrong bit value). At the same time, I assume that all counts in the main peak fall into the proper time bin and register as error-free key bits. To avoid being discovered, Eve needs to maintain the QBER at approximately the same level as before her attack. The premature counts are caused by non-zero optical power in the gap, which is caused in part by imperfect optical alignment between Eve and Bob. To estimate the required quality of optical alignment, I've measured time distributions at several values of $P_{\rm opt.~low}$ (at $P_{\rm opt.~high}={\rm 13~pW}$ and gap width of ${{\rm 2~\upmu s}}$). From the obtained data, I've calculated the probability ratio of having a premature click to having a click in the main response peak, for each used value of $P_{\rm opt.~low}$. The measurement has shown that this probability ratio rises approximately linearly with $P_{\rm opt.~low}$. If we additionally assume that this effect is the main contribution to the QBER and that Eve uses the intensity diagrams with power levels as given in Fig.~\ref{fig:faked-states-attack}(c), then the measurement data suggest
\begin{equation}
\label{eq:QBER-vs-misalignment}
\text{(QBER)}\approx \frac{4.2}{r_e},
\end{equation}
where $r_e$ is an extinction ratio between Bob's two detectors in the target basis achieved by Eve. Thus, to match values of the QBER in the 2--5\% range typically observed in QKD systems, Eve may need to achieve $r_e$ in the 19--23~dB range (or higher if other sources of errors are significant). This would be possible if the native extinction ratio of Bob's PBS exceeds $r_e$; this depends on the type of PBS used. Then, Eve would face a rather strong but probably realistic requirement on the precision of her polarization alignment. To narrow down the assumptions made in this assessment one would need to analyse and attack a concrete QKD implementation. This could be a task for the future.

\begin{figure}
\includegraphics[width=87mm]{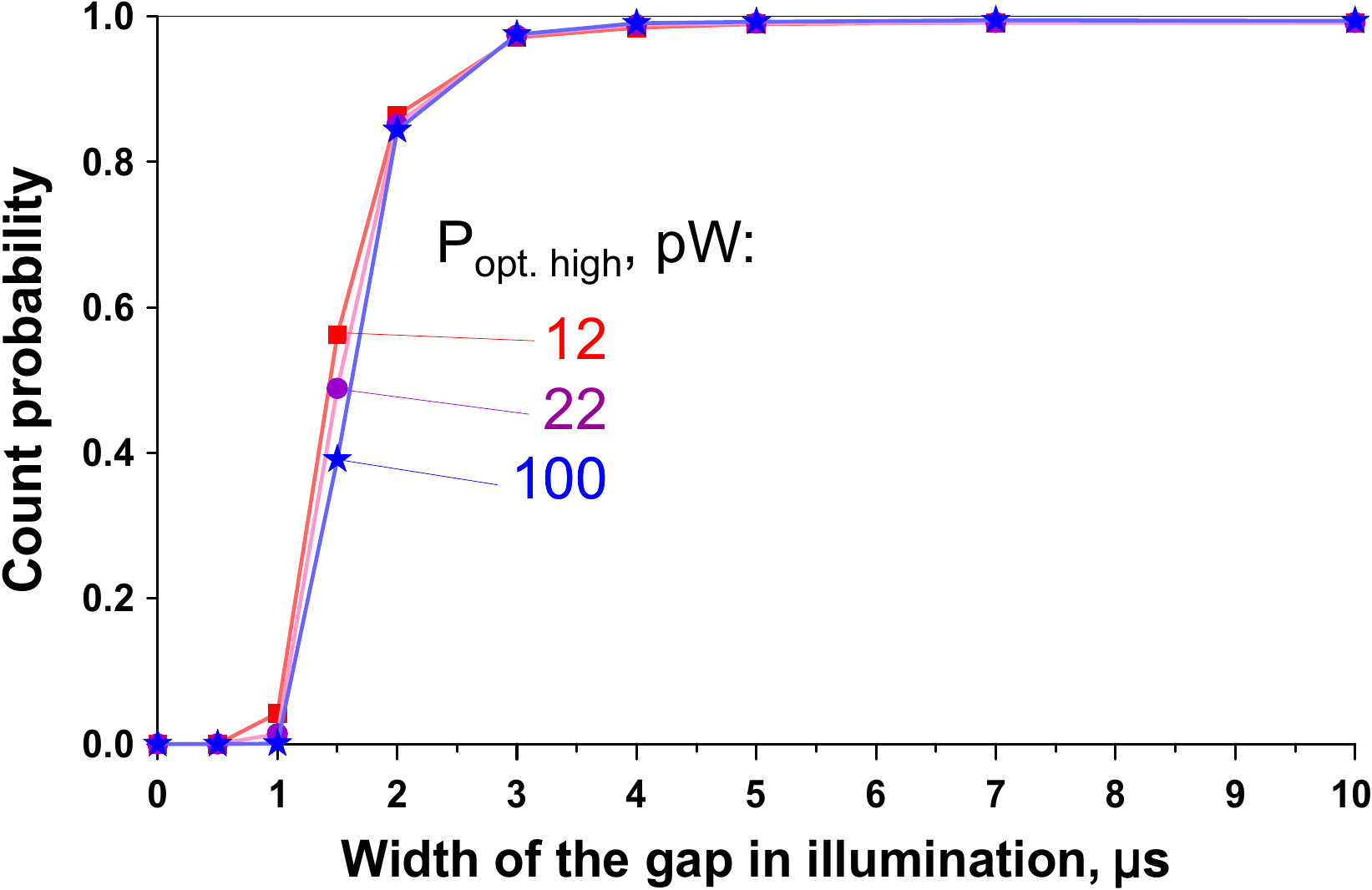}
\caption{\label{fig:p-wi}Detector model~1. Eve's probability of producing a pulse at the detector output vs. width of the gap in illumination (at $P_{\rm opt.~low}=0$).}
\end{figure}

Finally, Figure~\ref{fig:p-wi}
\begin{figure*}
\includegraphics[width=104mm]{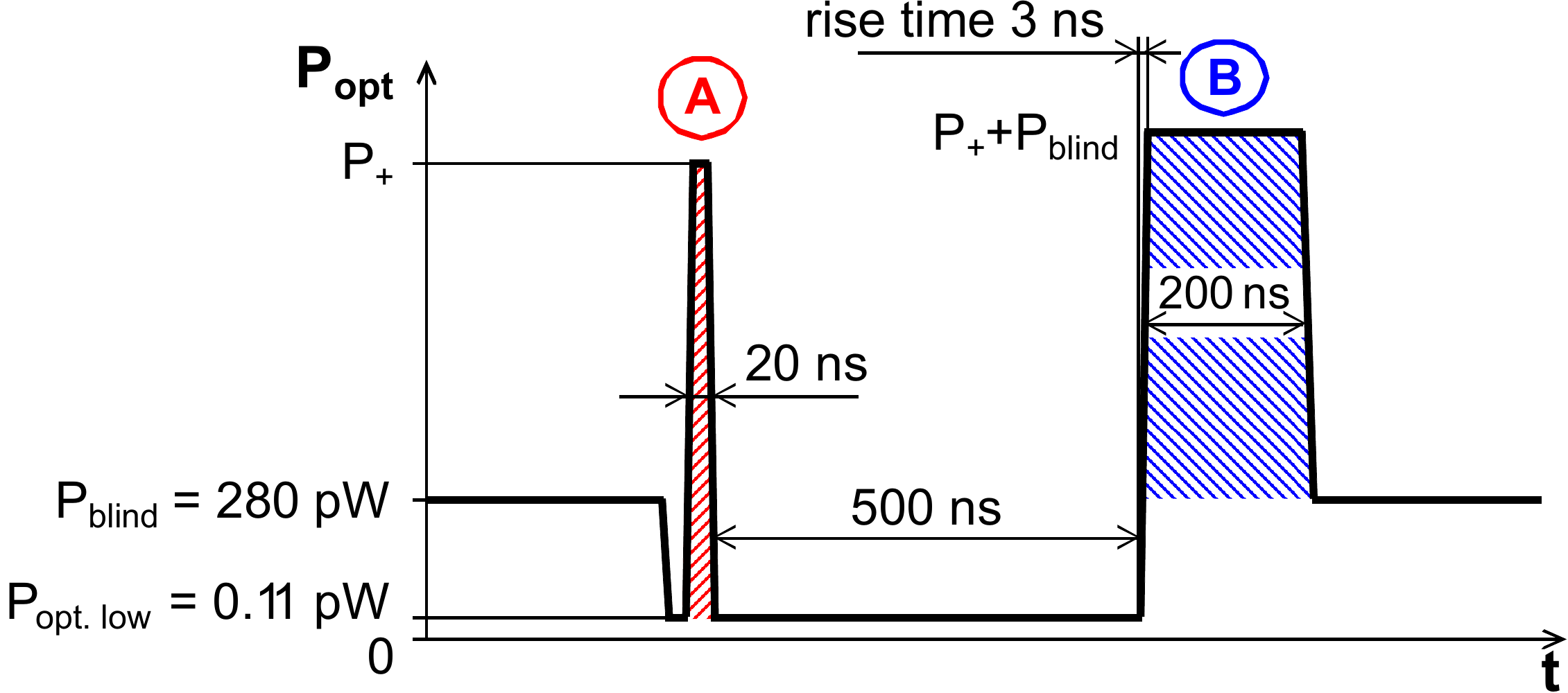}
\caption{\label{fig:detmodel2-control-diagram}Detector model~2. Control intensity diagram during testing. It was applied at 10~kHz repetition rate.}
\end{figure*}
shows how the probability of inducing the output pulse depends on the gap width. I have chosen the gap width of ${{\rm 2~\upmu s}}$ for all the other measurements with this detector, to achieve the count probability reasonably close to 1 without making the gap unnecessarily wide. As you can see, the count probability for ${{\rm 2~\upmu s}}$ or wider gap almost does not depend on $P_{\rm opt.~high}$. Interestingly, although the count probability exceeds 0.99 at gap widths larger than ${{\rm 5~\upmu s}}$, it never becomes exactly~1.

In these measurements, the gap repetition rate was 1~kHz. However, I have verified that, if necessary, the gap repetition rate can be increased to the limit. When the gaps follow each other in close succession (with less than ${{\rm 2~\upmu s}}$ between them), they still cause clicks at the SPD output.

\subsection{\label{sec:detector-model-2}Detector model~2}

This detector model is SPCM-200-PQ, industrially produced by EG\&G in the 1990s. While testing this model, I have focused on reducing the width of the main response peak. To achieve this goal, an improved control intensity diagram shown in Fig.~\ref{fig:detmodel2-control-diagram}
has been used. 780~nm illumination formed by mixing signals from two semiconductor lasers was applied uniformly over the entire photosensitive area of the APD 0.15--0.2~mm in diameter. The optical power at the APD is kept at the minimum blinding level $P_{\rm blind}={\rm 280~pW}$ most of the time. In the beginning of the gap, a short brighter pulse {\bf A} is applied. The purpose of this pulse is to discharge the capacitances in the SPD to about the same level every time in the beginning of the gap. Then the recharging process always starts at the same voltage and time, which leads to a certain voltage being applied to the APD at the end of the gap. If the pulse {\bf A} is absent, the recharge process starts at a random time of the last occurrence of avalanche before the gap. In this case, the APD voltage at the end of the gap varies, which leads to increased jitter in the single photon response \cite{ApplOpt-35-p1956}. The gap ends with another brighter pulse {\bf B}, which guarantees the arrival of the first few photons at the APD within a very short time. To fulfill this purpose, the pulse {\bf B} does not have to be long. However, the tested detector sample tended to produce double output pulses when $P_{\rm blind}$ was applied near the end of its first output pulse. Extending the length of the bright pulse {\bf B} to 200~ns reduced the probability of another output pulse appearing after the first one from 8\% to 0.5\%. In the 500~ns wide gap, illumination at the power level 34~dB below $P_{\rm blind}$ was applied to the detector, to simulate imperfect polarization splitting at Bob's PBS.

\begin{figure}
\includegraphics[width=87mm]{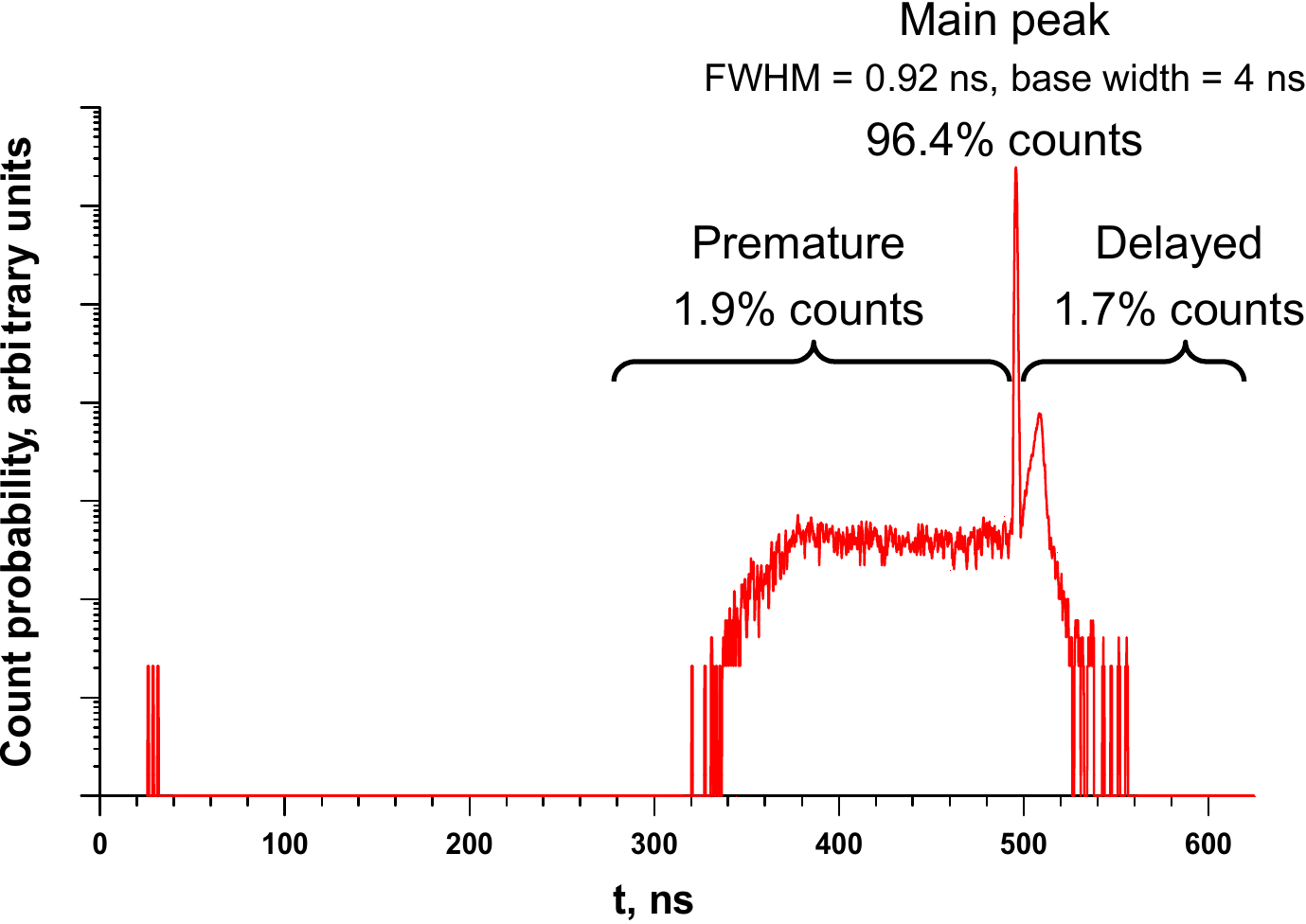}
\caption{\label{fig:detmodel2-timedistr}Detector model~2. Time distribution of pulse's leading edge at the detector output, when Eve controls it by an intensity diagram in which both parts {\bf A} and {\bf B} are present, at $P_+ = 784 \cdot P_{\rm blind} = 0.22~{\rm \upmu W}$.}
\end{figure}

The resulting time distribution of the SPD output pulses is shown in Fig.~\ref{fig:detmodel2-timedistr}.
At $0.22~{\rm \upmu W}$ peak power in the optical pulses {\bf A} and {\bf B}, the main response peak on the time distribution is 0.92~ns wide. The familiar premature counts in the gap are present on this time distribution, as well as delayed counts after the main peak. The latter can probably be attributed to small avalanches occurring early in the gap, resulting in delayed detector response after the end of the gap. The total probability, including the premature and delayed counts, of the detector producing a click in response to the control diagram is very close to~1.

\begin{figure}
\includegraphics[width=80mm]{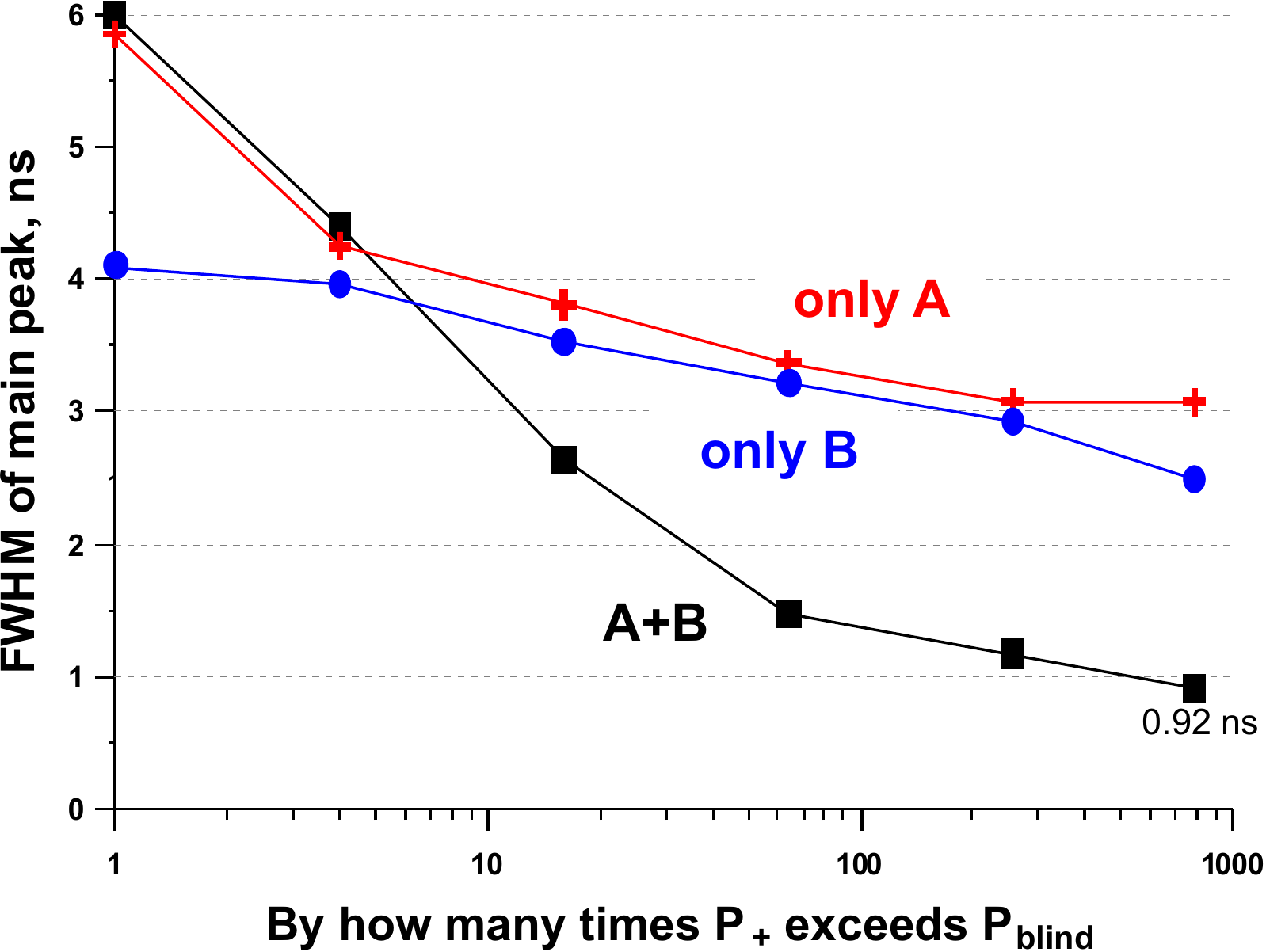}
\caption{\label{fig:detmodel2-fwhm-vs-p_plus}Detector model~2. Width of the main response peak vs.\ excess optical power $P_+$ in the parts {\bf A} and {\bf B} of the control intensity diagram. Three cases are charted: only part {\bf A} is present on the control diagram (while {\bf B} is not), only part {\bf B} is present (while {\bf A} is not), and both parts {\bf A} and {\bf B} are present simultaneously. Note that the leftmost point on the ``only~{\bf A}'' curve corresponds to roughly the same shape of the control diagram as was used for testing the detector model~1: a gap in constant illumination at the minimum blinding intensity.}
\end{figure}

The presence of both brighter optical pulses {\bf A} and {\bf B} on the control diagram is necessary to achieve the narrowest width of the main response peak. Figure~\ref{fig:detmodel2-fwhm-vs-p_plus}
shows how the width depends on the presence of each of these pulses and on the peak power in them.

\subsection{\label{sec:detector-model-3}Detector model~3}

This detector model is used in a compact passive polarization analyser module in Bob in a daylight free-space QKD system \cite{arXiv-0812-1880-quant-ph}. The system has been developed at the Centre for quantum technologies in Singapore. We have tested all four detector channels.
Unlike the previous two experiments, in this one we did not have physical access to measure absolute power impinging on the APDs. The saturation curves for the model~3 in Fig.~\ref{fig:saturc} are scaled based on a guess that the detector quantum efficiency in the linear part of the curves was around 50\%.

On this QKD system, Q.~Liu and myself have demonstrated that the Bob control method proposed in Sec.~\ref{sec:attack-against-QKD-system} works and that the detectors are individually addressable with sub-nanosecond jitter. We used polarization faked states that resulted in a control intensity diagram at the APDs similar to the one in Fig.~\ref{fig:detmodel2-control-diagram}. This will be reported in a separate article \cite{Qin-upcoming}.\\

From the experiments reported above, it appears that Eve might in practice be able to control passively-quenched detectors well enough to attack a real QKD system.

\section{\label{sec:countermeasures}Possibly affected systems and countermeasures}

Currently there are at least 28 papers reporting different QKD experiments that employ non-gated Si APDs.
These papers break down as follows.
Eight of them reported the use of passively-quenched APDs \cite{NewJPhys-4-p43,OptExpress-12-p3865,TechPhys-50-p727,NewJPhys-8-p249,PhysRevLett-98-p010504,ApplPhysLett-89-p101122,arXiv-0812-1880-quant-ph,PhysRevA-78-020301R},
ten reported the use of non-gated, actively quenched APDs \cite{PhysRevA-57-2379,PhysRevLett-81-p3283,PhysRevLett-84-p5652,OptExpress-13-p202,NewJPhys-8-p32,ApplPhysLett-89-p191121,IEEEJQuantumElectron-40-p900,OptExpress-13-p3015,IEEEJQuantumElectron-43-p130,OptExpress-16-16840},
and another ten did not specify the type of quenching, only saying Si APDs or ``detectors'' (which I assume were Si APDs) were used \cite{OptLett-21-p1854,ElectronicsLett-37-p512,Nature-420-p762,PhysRevLett-89-p187901,NewJPhys-6-p92,Nature-419-p450-and-ProcSPIE-4917-p25,OptExpress-12-p2011,PhysRevLett-94-p150501,ProcQuantumElectronicsConference-2005-EQEC05-305,NaturePhysics-adv-nphys629}. I have since learned that three of the latter ten experiments \cite{NaturePhysics-adv-nphys629,Nature-419-p450-and-ProcSPIE-4917-p25,ProcQuantumElectronicsConference-2005-EQEC05-305} did in fact use passively-quenched detectors of a design very similar to the models~1 and 3 studied in this paper. Thus, it appears that passively-quenched and actively-quenched Si APDs are equally frequently used in QKD experiments. I remark that at least one model of actively-quenched Si SPD has been shown vulnerable to a somewhat similar attack also involving bright illumination \cite{arXiv-0809-3408-quant-ph}.

Continued frequent use of passive quenching can be explained by its practical properties. It is well known that an actively-quenched APD delivers superior count rate and timing characteristics \cite{JModOpt-51-1267,ApplOpt-35-p1956}. However, a passively-quenched circuit is simpler, cheaper, and more versatile; the biasing parameters are easy to adjust; a larger photosensitive area APD can be used than those embedded in commercially available actively-quenched detector modules. At the same time, the performance of the passively-quenched SPD is often adequate for the task. For example, in the 144~km QKD experiments \cite{PhysRevLett-98-p010504,NaturePhysics-adv-nphys629}, laboratory-made passively-quenched detectors were used because the average count rate at Bob was low \cite{144km-passive-quenching-adequate}. 
 
Unfortunately, none of the 28 experiments in my literature sampling seemed to implement any countermeasure against bright-light attacks (with the possible exception of Ref.~\onlinecite{PhysRevLett-84-p5652} where Eve's illumination might accidentally cause Bob's separate timing detector to work incorrectly). Neither do I know of any SPD module with a specified guaranteed behavor under bright-light illumination, or equipped with an extra output that signals saturation or blinding.

It may appear that introducing authenticated timing into the QKD protocol can prevent my attack. However, Eve can try a slightly different tactics discussed in Sec.~\ref{sec:attack-against-QKD-system}, by starting to form a faked state before the actual detection occurs in Bob$^\prime$. This tactics may in practice allow her to mimic the timing of Bob's clicks with just a few ns extra delay. Additionally, when the QKD system uses optical fiber \cite{OptExpress-12-p3865}, Eve can gain time by routing her classical communication from Bob$^\prime$ to the FS (see Fig.~\ref{fig:faked-states-attack}(b)) via a radio link in which signals propagate faster than in the fiber. Authenticated timing does not prevent the detector controllability, and thus is not a complete solution.

Other researchers have proposed to equip each SPD with a ``detector ready'' signal that is only present when the voltage at the APD guarantees certain minimum quantum efficiency \cite{hack-proofing-munich}. I think, this is a promising idea. These ``detector ready'' signals from all Bob's SPDs can be combined on an AND-gate and used to disable/enable click recording from all SPDs simultaneously by Bob's electronic registration system. Besides preventing the bright-light attacks, this would also be useful to thwart subtlier exploits. This circuit introduces registration blanking time for all detectors simultaneously whenever at least one of them is insensitive to photons after an avalanche. Rejecting clicks that occur whenever at least one detector is having a deadtime seems to be a necessary security measure in any QKD system \cite{hack-proofing-munich}. Additionally, this photon registration system can guarantee a certain quantified minimum quantum efficiency of each detector whenever the system is recording clicks. This guarantee may be required by a general security proof that takes into account equipment imperfections \cite{minimum-eta-proof}.

Once a hack-proofed system is built, it would have to be tested thoroughly under bright-light illumination with various temporal diagrams over a wide input intensity range. Ideally, the testing should include higher input power levels up to and above the damage threshold of Bob's optics.

\section{\label{sec:conclusion}Conclusion}

In this paper, I have shown how the saturation and blinding behavior of the passively-quenched APD can be used to gain control over detectors and stage an attack against a QKD system. Passively-quenched detectors of three different models have been experimentally tested and their control demonstrated by the same method, under realistic conditions. It would now be interesting to demonstrate a complete attack against a running QKD system.

\begin{acknowledgments}
The author is grateful for hospitality, assistance, useful discussions and a loan of equipment to Sergei Kulik and his colleagues at the Laboratory of spontaneous parametric down-conversion at the Moscow State University; to Oleg Kotov and his colleagues at the Fiber optics laboratory at the St.~Petersburg State Polytechnical University; to Alexander Ushakov; to Christian Kurtsiefer, Ilja Gerhardt and colleagues at the Centre for quantum technologies in Singapore. The author thanks the anonymous reviewer for valuable critique of the first version of this paper. Johannes Skaar and Lars Lydersen are thanked for comments on the manuscript. Financial support from NTNU, the Research Council of Norway (grant no.\ 180439/V30), and
Korea Science and Engineering Foundation 
(grant no.\ R01-2006-000-10354-0) is acknowledged. This work was also partially supported by the National Research Foundation \& Ministry of Education, Singapore.
\end{acknowledgments}


\bibliography{paper}

\end{document}